# Experiencing avatar direction in low cost theatrical mixed reality setup

**Georges Gagneré**
University Paris 8
2 avenue de La Liberté 93326 Saint-Denis
georges.gagnere@univ-paris8.fr

**Cédric Plessiet**
University Paris 8
2 avenue de La Liberté 93326 Saint-Denis
cedric.plessiet@univ-paris8.fr

## ABSTRACT

We introduce[1] the setup and programming framework of AvatarStaging theatrical mixed reality experiment. We focus on a configuration addressing movement issues between physical and 3D digital spaces from performers and directors' points of view. We propose 3 practical exercises.

## CCS CONCEPTS

• **Human-centered-computing** → **Interaction design** → Empirical studies in interaction design; • **Computer systems organization** → **Real-time systems** → Real time system architecture; • **Applied computing**→ **Arts and humanities**→ performing arts • **Computing methodologies** → **Computer graphics** → **Animation** → Motion capture

## KEYWORDS

Augmented acting, avatar direction, mixed reality, motion capture, performing arts

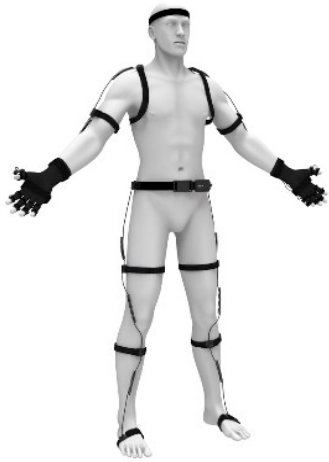

**Figure 1: Perception Neuron motion capture suit**

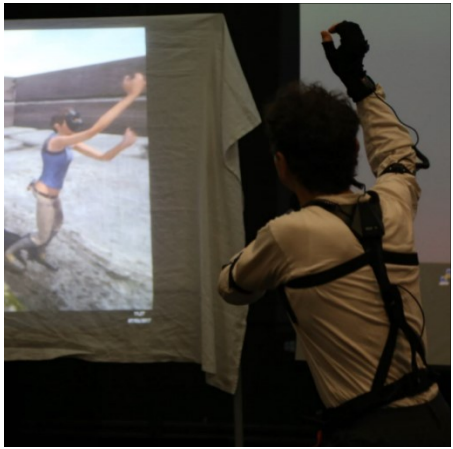

**Figure 2: Mocaptor controlling her avatar**

## 1 INTRODUCTION

We are sharing the results of a theatrical experiment based on relationships between performers and avatars in an environment mixing 3D digital space and physical « traditional » stage space. We organize the way in which two states of reality (the virtual world of 3D digital assets and the performer's physical world) work together for the pleasure of an audience.

We focus in this paper on avatar acting issues and technological solutions we experienced to conduct theatrical artistic processes all along 2017, that resulted in public performances in December 2017 at Le Cube, digital creative center in Issy-les-Moulineaux, near Paris [6]. And we propose 3 practical exercises to enact the theoretical issues.

### 1.1 On the physical side of the mixed reality setup

Our research is rooted in the confrontation of a theatrical acting process and an avatar real-time motion capture control process on the same physical stage. We started by using a Kinect device, and we went on using one of the low-cost geo-spatial motion capture suits that appeared on the market around 2015 (fig. 1) [11]. The performer wearing the mocap suit is called the 'mocaptor' (fig.2) – and we simply called the physical actor without a mocap suit in front of the avatar 3D screen the 'performer'. We'll therefore have onstage mocaptors, avatars and performers.

At its inception in 2014, our research explored two hypotheses: on one hand, we would offer possibilities for the stage director and the audience to access the digital nature of 3D scenery and acting avatars by wearing HMD helmets; on the other hand, we would provide access to the digital space by projecting it onto a fixed screen at the rear of the stage [3] [4].

Because of the theatrical context of the 2017 experience, we adopted the second hypothesis. We allowed all the collaborators (performers, mocaptors, stage director, digital artist), but also the physical spectators, who are the fundamental part of the theatre live process, to share together the physical side of the mixed reality set up.

### 1.2 Keeping a visual feedback to control the avatar acting process

Our choice to work without HMD consequently prevents the mocaptor from accessing her digital body from an immersed point of view. She only can see her digital body through the 2D screen on the stage, or through « feedback » monitors placed downstage, which allow for rather more distant conditions of body awareness. Nevertheless, the acting issues from the immersive point of view [1] [6], raised by the fact of inhabiting an avatar 'second body' and making her present to others (performer or audience), remains valuable.

This situation is familiar in theatre through the figure of a puppeteer manipulating a puppet and maintaining a distant point of view on it. Wearing a mocap suit puts the mocaptor in the same situation, overseeing a 'natural' transfer of body movements. And we decided to keep her intimately involved in the control of her avatar by giving her a permanent visual feedback.

## 2 MIXED REALITY EXPERIMENT – FROM MOCAPTOR TO AVATAR

### 2.1 Spaces in hybrid interrelation – AvatarStaging framework

The experimental setup mixes 5 spaces and is based on AvatarStaging framework [9] [10] (fig. 3):
- Space A: physical stage
- Space B: 3D digital scenery, projected on a 2D screen
- Space C: mocaptor space
- Space D: space where digital artist and operator works out the 3D avatars and scenery
- Space E: audience and stage director space

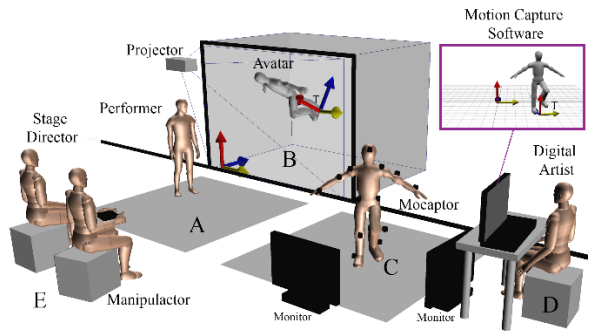

**Figure 3: AvatarStaging mixed reality setup**

The mixed reality stage results from interrelations between spaces A and B. The avatar's movements in B and the relationships between A and B depend upon the action of the mocaptor in C and the real-time play-out and operation in D. The resulting image of space B (on the video projection screen) has the same appearance as a 2D movie image shot with a camera. Nevertheless, it is fundamental to keep in mind that B is a (virtual) 3D space, similar to the physical stage in front of the video screen.

### 2.2 From the mocaptor to the avatar

We take from computer graphics conventions a way of describing the movement of a character [2]. The mocaptor performs relative movements with her body parts (legs, arms, head, etc.), in relation to a point called her 'reference' T (a starting point within the digital spatial grid).

As we work in a digital space B, we need to independently set the reference T of the digital character, built by the software associated with the motion capture suit, in the coordinates system of B. This operation can completely change the position of the avatar in relation to the position of the mocaptor. Firstly, this fundamental separation between digital space B and physical space C is bodily difficult to deal with for a performer used to acting on a traditional stable stage. Secondly, we can control in real time the parameters of the avatar's T, that is deeply disorienting from the perspective of the mocaptor.

Thinking about the 3D nature of avatar space B, we first imagine an approximate analogical reproduction of the digital scenery in mocaptor space C in order to make the avatar movement easier for the mocaptor to perform. But it quickly became too difficult to implement because of multiplication of feedback constraints.

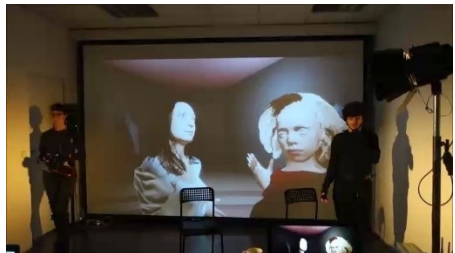

**Figure 4: Reharsal**

### 2.3 Reducing the mocaptor space C

Relying on the separation between spaces B and C, we reduce the scope of acting space C inside an absolute volume, that would be projected in space B relative to the digital scenery. We considered this smaller absolute volume in facilitating the establishment of the mocaptor's visual contact with the avatar through feedback monitors (fig. 4).

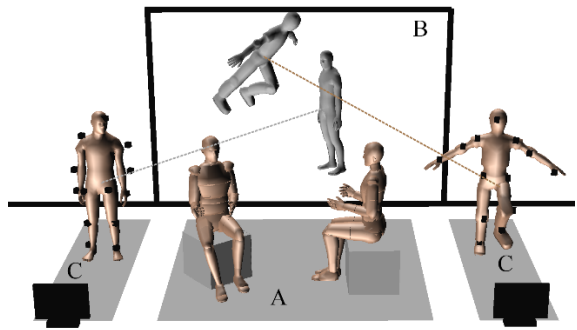

**Figure 5. Configuration of AvatarStaging**

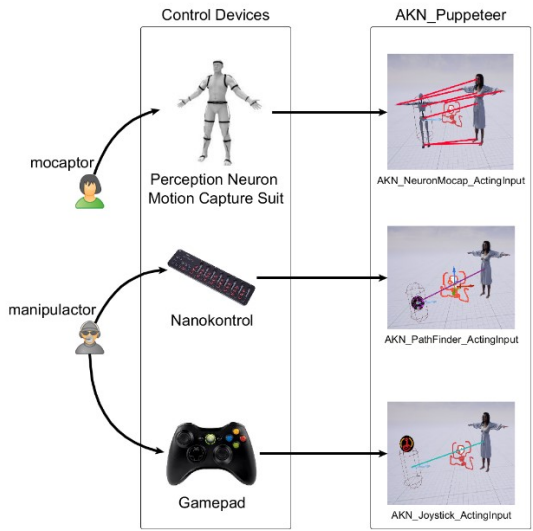

**Figure 6. Programming framework AKeNe**

Figure 5 shows one basic configuration we used during the 2017 artistic and pedagogical experiments. This setup will also be used for the 3 exercises. It consists in two corridors on each lateral side of the stage A with a feedback monitor on the downstage extremities. The mocaptors use the big video projection screen as an alternative feedback when they are walking upstage. This set-up optimizes the visual feedback arrangement, given that the mocaptors largely kept their head forward facing along the corridor.

In this configuration the mocaptors are very close to the performers and keep an indirect but strong connection with the physical stage. It helps in sharing the same acting energy. The director can compare the mocaptors' movements and the results of their transposition on the avatars' bodies, so as to focus on efficient directorial responses with a view to enhancing expressivity.

## 3 MANIPULACTOR ASSISTANCE FOR CONTROLLING THE AVATAR

### 3.1 The manipulator, a puppeteer of the mocaptor

Regarding the scenic actions on the physical stage A, we know that performers often act without keeping an absolute reference to the space. When they are in dialogue with each other, they often take care to offer their face to the audience. In doing so, they often must cheat on the effective line of sight of their partner, because of lighting positions or the orientation of scenery. For opera singers, the imperative need to project the voice towards the audience asks them to deal with 'artificial' spatial relations. They also must permanently keep in touch with the conductor through multiple video feedbacks all around the stage.

Playing with the parameters of the avatar reference transform T (see 2.2) allows one to orient the reduced acting volume of the mocaptor in relation to scenic needs. We can use the metaphor of a puppeteer guiding the puppet through the stage, the difference being that the extent of the manipulation is not limited by physical constraints while we are in a digital and real-time-rendered space. We give the responsibility to guide the avatar to an assistant (of both the digital artist and the stage director), who should both understand acting rules and computing possibilities: we called him the 'manipulator'.

### 3.2 AKeNe programming framework

The programming framework allowing the AvatarStaging experiments uses the AKeNe library [9]. The development is focusing on two main directions. On one side, we use and improve modules for combining movement data and solving motion retargeting issues. For instance, we use a puppeteering system that allows extraction of the mocap data to mix them with inputs coming from a gamepad, or even from AI videogame modules as pathfinder (fig. 6).

On the other side, we use an event manager for cueing positions and actions in the digital scenery.

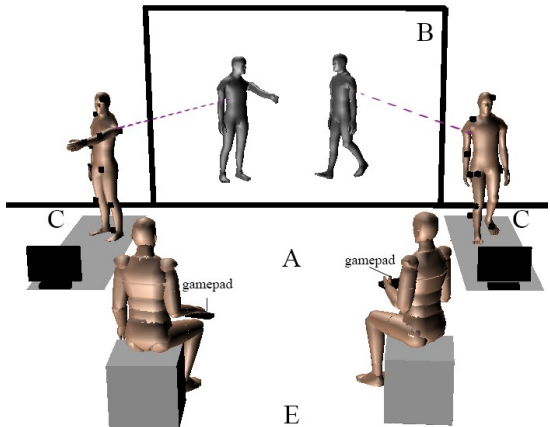

**Figure 7. Situation 1 – Walking**

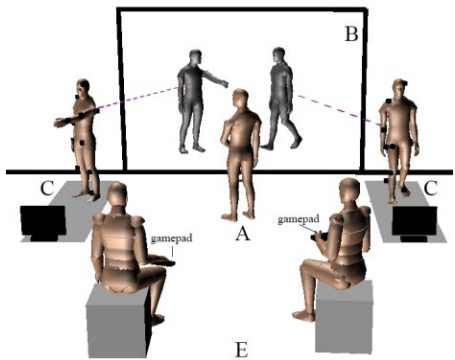

**Figure 8. Situation 2 – Walking and watching**

### 3.3 Actions and tool of the manipulactor

The manipulactor is finally able to guide the avatar by:
- moving in a forward/backward direction
- moving laterally to the right or the left
- moving up and down
- rotating the yaw for changing her forward direction
- rotating the pitch to put her on a horizontal position and make her float in the air

All these actions can be done in addition to the mocaptor movements. We use a gamepad controller for applying them to the avatar, taking advantage of the powerful design of the device for combining and fine-tuning the appropriate transformations.

The manipulactor focuses on two main goals:
- adjusting the scenic address of the avatar towards its performing partner in the A space, from the point of view of the audience; and respecting perspectival constraints [5]
- accompanying the mocaptor in augmented movements that are usually not accessible to a performer, as, for instance, floating in the air.

## 4 PRACTICE: 3 EXERCICES

We finally describe three practical situations aimed both for demonstrating AvatarStaging setup and programming framework AKeNe, and for exercising as a mocaptor or manipulactor to achieve expressive avatar movements.

### 4.1 Situation 1: Walking

The first situation (fig. 7) consists in having avatars naturally walking and pausing in the digital space. Mocaptors and manipulactors shall collaborate to expand the constrained walking space in C to the larger space B, trying to keep a natural pacing. They will also be asked to propose basic acting interactions between the avatars, as interlacing trajectories or looking at each other.

### 4.2 Situation 2: Walking and watching

From the previous situation, we add a performer in front of the screen as a third partner (fig. 8). We'll explore the complexity of changing watching goal from the other avatar to the performer, and vice versa.

### 4.4 Situation 3: The crowd

We finally propose to improvise with a group of 6 avatars in motion (fig. 9). Each mocaptor controls 3 avatars (identical or not), and each avatar is guided by one manipulactor.

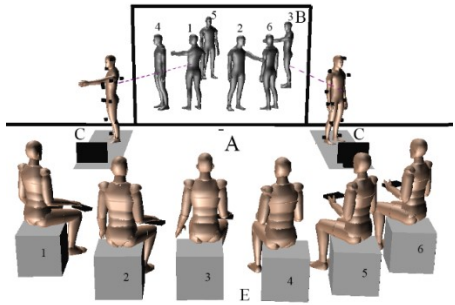

**Figure 9. Situation 3 - The Crowd**

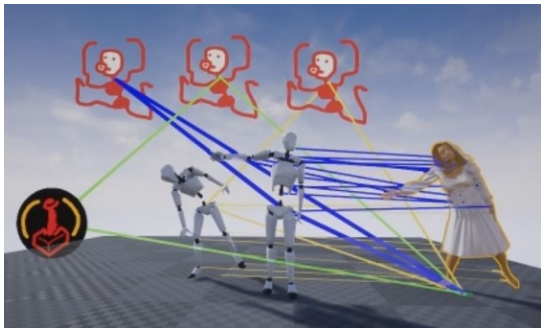

**Figure 10. Complex puppeteering**

## 5   PERSPECTIVES

In our experiments, we felt that the mocaptor is a kind of puppeteer inside the avatar, where the avatar has a "physiology" close to that of a wooden puppet skeleton (or a stylized robot). When the avatar is a 3D scan with imperfect skinning, the mocaptor has the feeling of living inside a kind of mascot - that's to say, a body mask (or human-sized doll).

The action of the manipulactor is like that of the traditional puppeteer who is external to the puppet. One could then compare the relationship between mocaptor and manipulactor to that (extraordinarily) of a living puppet who would suddenly see some of her wires taken back by an outside hand. We therefore need to establish avatar control-sharing protocols for setting up specific actions (scenic address, large space routes, stage entries).

When the mocaptor (both puppeteer and puppet), walks under the manipulactor's control, she finds herself in the position of a golem, a mythological creature capable of movement but devoid of free will. We forecast to extend the mocaptor/manipulactor relationships by exploring other movement combinations and plausible movement qualities. Hybridisation with autonomous virtual actors is one direction [8]. Addressing facial expression and splitting control in between several mocaptors is another (fig. 10).

## REFERENCES

1   Amato A., Perény E. (dir.), *Les avatars jouables des mondes numériques. Théories, terrains et témoignages de pratiques interactives*, Ed. Hermes Lavoisier, 2013
2   Foley J. D., Van Dam A., Feiner S. K., Hughes J., *Computer graphics. Principles and practices,* Ed. Addison Wesley, 3rd edition, 2013
3   Gagneré G., Plessiet C., « Traversées des frontières » in Frontières numériques & artéfacts (sous la direction de Hakim Hachour, Naserddine Bouhaï & Imad Saleh), L'Harmattan, 2015, Chapitre 1, pp. 9-35
4   Gagneré G., Plessiet C., « Perceptions (théâtrales) de l'augmentation numérique » in Frontières numériques & artéfacts (sous la direction de Hakim Hachour, Naserddine Bouhaï & Imad Saleh) , actes du colloque international Frontières Numériques : Perceptions, Toulon, décembre 2016.
5   Gagneré G., Plessiet C., Sohier R., «  Espace virtuel interconnecté et Théâtre. Une recherche-création sur l'espace de jeu théâtral à l'ère du réseau », in Revue : Internet des objets , Num. 2, Vol. 2, juin 2017, ISTE OpenScience
6   Gagneré G., Plessiet P., A. Lavender, T. White, « Challenges of movement quality using motion capture in theater», in *Proceedings of ACM MOCO conference, Genova, Italy, June 2018*
7   Morie, J.F., 'Performing in (virtual) spaces: Embodiment and being in virtual environments', International Journal of Performance Arts and digital Media 3: 2&3, 2007, pp. 123–138
8   Plessiet, C., S. Chaabane, and G. Khemiri. "Autonomous and Interactive Virtual Actor, Cooperative Virtual Environment for Immersive Previsualisation Tool Oriented to Movies." In *Proceedings of the 2015 Virtual Reality International Conference,* 5:1–5:4. VRIC '15. New York, NY, USA: ACM, 2015.
9   Plessiet C., Gagneré G., Sohier R., « Avatar Staging: an evolution of a real time framework for theater based on an on-set previz technology », in *Proceedings of the 2018 Virtual Reality International Conference*. New York, NY, USA: ACM, 2018
10  www.avatarstaging.eu
11  https://neuronmocap.com